# A 4.9-GHz Low Power, Low Jitter, LC Phase Locked Loop

**Tiankuan Liu**[*]

Southern Methodist University,
Dallas, Texas 75275, USA
E-mail: liu@physics.smu.edu

ABSTRACT: This paper presents a low power, low jitter LC phase locked loop (PLL) which has been designed and fabricated in a commercial 0.25-μm Silicon-on-Sapphire CMOS technology. Random jitter and deterministic jitter of the PLL are 1.3 ps and 7.5 ps, respectively. The measured tuning range, from 4.6 to 5.0 GHz, is narrower than the expected one, from 3.8 to 5.0 GHz. The narrow tuning range issue has been investigated and traced to the first stage of the divider chain. The power consumption at the central frequency is 111 mW.

# Contents

## 1. Introduction

The upgrade from Large Hadron Collider (LHC) to super-LHC imposes new challenges on the design of ATLAS Liquid Argon Calorimeter readout system. As a key part of the readout system, the optical data links in the proposed upgrade scheme must operate at the data rate of about 100 gigabit per second (Gbps) per front–end board (FEB), 60 times higher than the present data rate, with the same power consumption as that of the present system [1]. The serializers used in the present optical data link system cannot meet the upgrade requirements on data rate and power consumption. No commercial serializer meets the requirements on radiation tolerance. For the upgrade of the optical data links, we have successfully designed a radiation tolerant, high speed, and low power 16:1 serializer Application-Specific Integrated Circuit (ASIC) operating at 5 Gbps with a ring oscillator based phase locked loop (PLL) [2]. In order to achieve higher data rate and density than the present prototype, we aim at developing a multi-channel serializer array with 8–10 Gbps per channel. Correspondingly, we have to develop a PLL operating at 4–5 GHz. The ring oscillator based PLL cannot be used due to its limits on speed, jitter, and power consumption. A high speed, low jitter, and low power LC-tank based PLL (LCPLL) has been designed, fabricated and tested. The design of the PLL has been presented before [3]. The test results are presented in this paper.

## 2. Design

Since the design has been presented before, only a brief summary is described here for context information. The block diagram of the PLL is shown in Figure 1. An LVDS receiver and a CML driver are added as the input and output interface for test purpose. In the block diagram, the PFD is a phase and frequency detector. The charge pump (CP) converts the up and down signals into control current. The low pass filter (LPF) integrates the current into control voltage. The VCO is a LC-tank-based voltage controlled oscillator (VCO). The divider chain consists of four divide-by-2 dividers, a high speed CML divider and 3 CMOS dividers. Because the bandwidth of the CML driver is not high enough, we monitor the output of the CML divider rather than the output of VCO.

The ASIC was submitted for fabrication in August 2009 and delivered in November 2009. The die micrograph of the PLL is shown in Figure 2. The PLL is located at a corner of a 3 ×3 mm$^2$ chip area shared by the PLL, a serializer and other small test structures. The PLL itself is 1.4 ×1.7 mm$^2$. The die micrograph is shown in Figure 2.

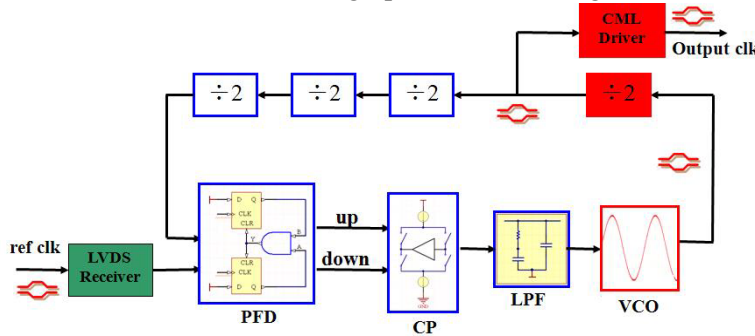
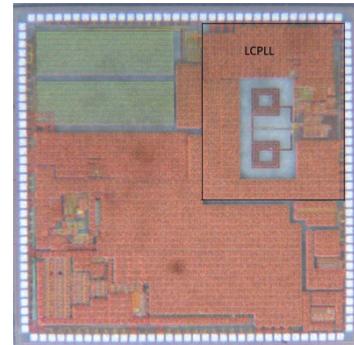

Figure 1. Block diagram of the LCPLL

Figure 2. Die micrograph

## 3. Test results

### 3.1 Test setup

The ASIC is wire bonded to a printed circuit board (PCB). The input reference clock and the output clock signals are transmitted through microstrips and SMA connectors. The input clock is frequency adjustable square wave signals generated by a pattern generator (Model MP1763C produced by Anritsu Company). The output clock is monitored using a real-time oscilloscope (Model TSA72004 produced by Tektronix, Inc.) with a differential probe (Model P7380SMA). The filter bandwidth in the PLL is set at 2.5 MHz and the charge pump current is set at 20 µA. The waveforms of the PLL output clock locked to the input reference clock are shown in Figure 3.

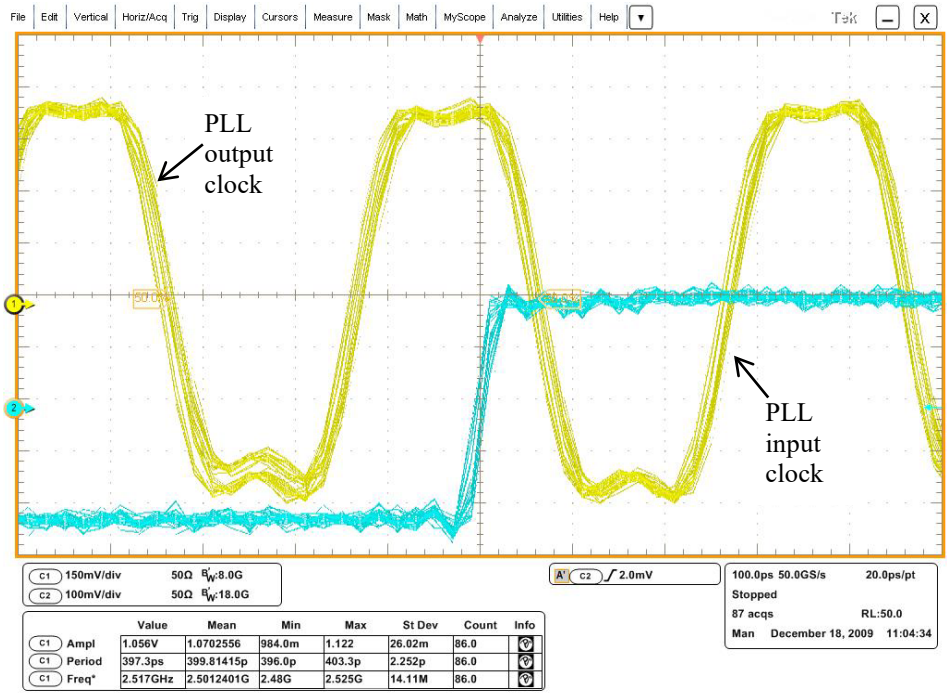

Figure 3. Waveforms of the PLL output clock locked to the input clock

Jitter of the PLL output clock is analyzed using commercial jitter analysis software (Tektronix TDSJIT3). An example of jitter analysis results is shown in Figure 4. Deterministic jitter is contributed only from periodic jitter. Data dependent jitter is zero because the signal to be analyzed, which is a clock, has only a single pattern. Duty cycle jitter is zero because jitter for a clock is analyzed only at rising edges.

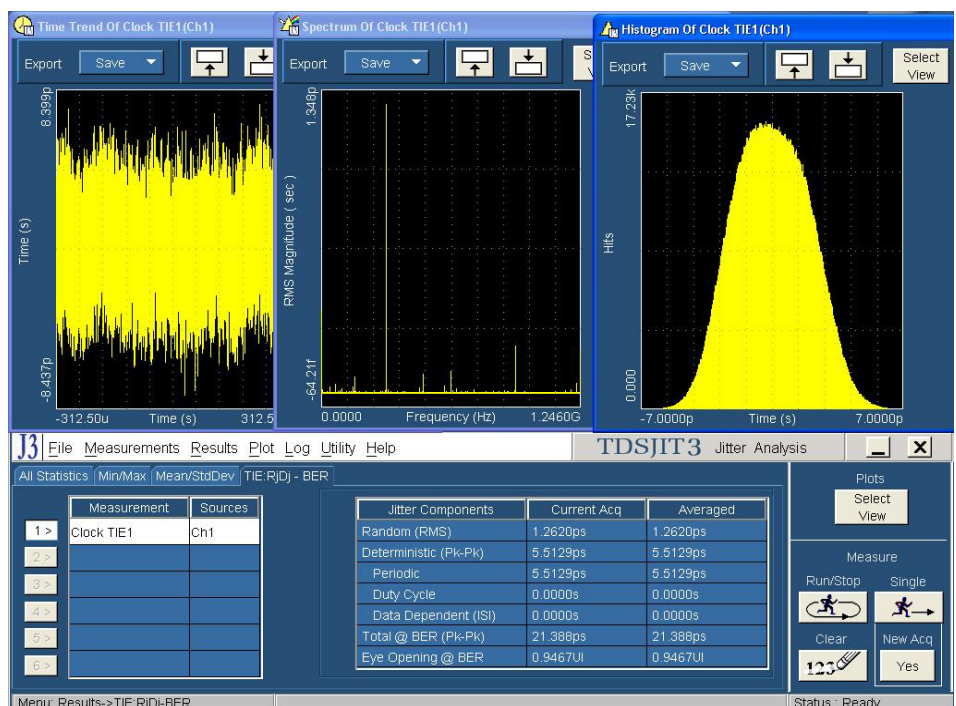

Figure 4. Jitter analysis results of the PLL output clock

### 3.2 Test results

Seven boards have been measured. Two boards do not work because their power supplies are either open or short, which we are investigating. The test results of the other five working boards are shown in Table 1. The measured tuning range is from 4.6 to 5.0 GHz, narrower than the expected value of from 3.8 to 5.0 GHz at the typical corner. The narrow tuning range issue has been investigated and will be discussed in the next section. The power consumption at the central frequency of each PLL is 111 mW, excluding that consumed by the CML driver and the LVDS receiver. The power consumption of an LCPLL is lower than that (173 mW at 2.5 GHz) of the ring oscillator based PLL used in the 16:1 serializer. Random jitter of 1.3 ps (RMS) is larger than the value estimated from the phase noise of the VCO, suggesting that the contribution from the other parts of the PLL cannot be ignored. Deterministic jitter of 7.5 ps (peak-peak) is larger than expected and will be investigated.

| Tuning frequency (GHz) | Upper | 4.98±0.02 |
|---|---|---|
| | Lower | 4.63±0.12 |
| At the central frequency | Power consumption (mW) | 111±8 |
| | Amplitude (pk-pk of differential, pk-pk, V) | 1.23±0.09 |
| | Rise time (20% - 80%, ps) | 44.9±2.4 |

| | Fall time (20% - 80%, ps) | 44.4±2.2 |
|---|---|---|
| | Random jitter (RMS, ps) | 1.3±0.3 |
| | Deterministic jitter (pk-pk, ps) | 7.5±1.1 |

Table 1. Measured average values and standard deviation of five LCPLLs

### 3.3 Investigation of narrow tuning range

In order to investigate the narrow tuning range issue, we monitor the PLL control voltage when we change the input reference clock frequency. The PLL control voltage node is connected to a die pad through a transmission gate. When the PLL input clock frequency decreases from 311 MHz to 297 MHz, the output clock frequency which is half of the VCO output frequency decreases from 2.49 GHz to 2.38 GHz and the control voltage decreases from 2.4 V to 1.6 V as expected. However, when the PLL input clock frequency is lower than 297 MHz, the control voltage drops to 0.1 V quickly and the output clock signal is stuck at 2.83 GHz. The measured PLL output clock frequency versus control voltage (V-F) curve is shown in Figure 5. The voltage-versus-frequency curve is not monotonic as expected. The measurement errors are also shown in the figure. The measurement errors are large when the control voltage is less than 1.6 V, indicating that the PLL output clock waveforms are unstable.

One possible reason of the narrow tuning range is the inaccuracy of varactor model since the V-F curve of a VCO is directly related to the voltage-versus-capacitance (V-C) curve of varactors used inside a VCO. A varactor is not a standard supported device in the technology we use. The varactor used in the design is a regular threshold voltage N-channel (RN) MOSFET with its source and drain terminals connected together. A separate varactor is implemented in the same die to evaluate the varactor model. The measured and simulated capacitance-voltage curves of a varactor are shown in Figure 6. Measured C-V curve of the varactors demonstrates that a MOSFET-based varactor behaves as expected, though the measured values deviate from the simulated values, which will be investigated in future. The measurement results rule out the suspicion that inaccuracy of varactor model causes narrow tuning range.

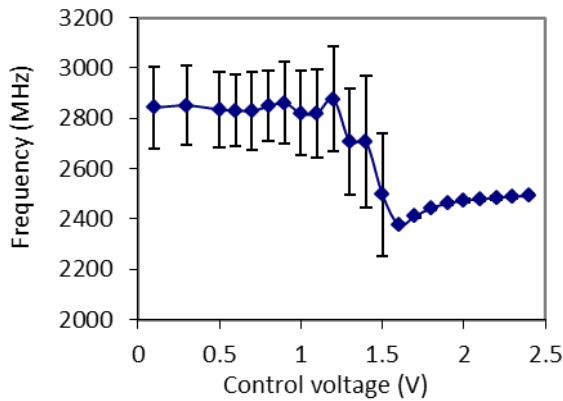

Figure 5. The measured PLL output frequency versus control voltage

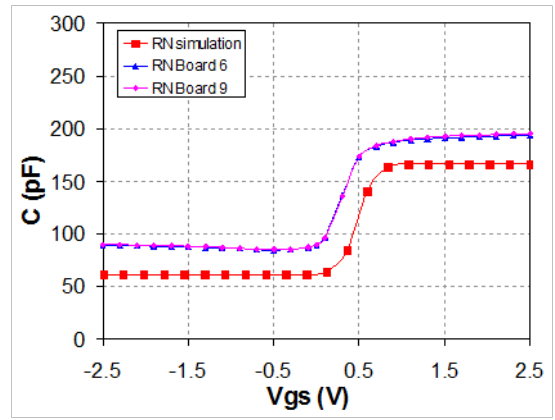

Figure 6. The measured and simulated capacitance-voltage curves of a varactor

We compare the observed waveforms of the PLL output clock with the simulated waveforms and conclude that the first stage of the divider chain causes narrow tuning range. We force the control voltage to 1.17 V by connecting the control voltage node to an external voltage source and monitor the output clock waveforms, which are shown in Figure 7(a). The output waveforms exhibit non-uniform cycles, which is the reason of the large measurement errors in the frequency measurements shown in Figure 5. The simulated waveforms at the fastest active devices with the possibly slowest passive devices corner, which is closest to that indicated by the fabrication process sample devices, are shown in Figures 7(b). The output waveforms of the VCO are still uniform in both amplitude and period, whereas the output waveforms of the CML divider are not. Note that the frequency of the CML divider output waveform is higher than one half of that of the VCO output waveform. In other words, the CML divide-by-2 divider does not work as expected.

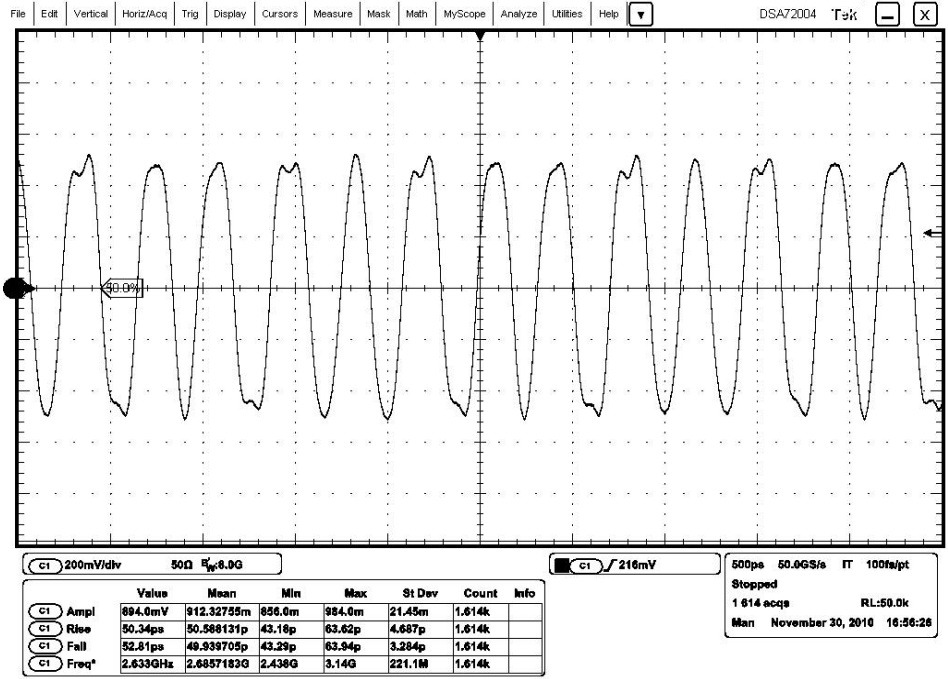

Figure 7(a). The output waveform when the control voltage is 1.5 V

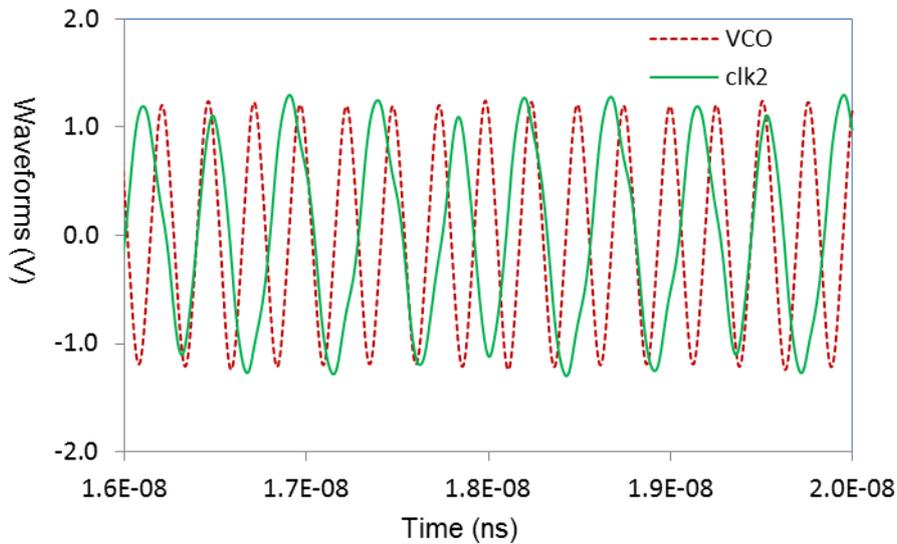

Figure 7(b). Simulated output waveforms of the VCO (red dashed) and of the CML divider (green solid)

The reason of the abnormal cycles is understood as follows. The schematic of the CML divider is shown in Figures 8(a). The CML divider consists of a master latch and a slave latch. The schematic of the latch is shown in Figure 8(b). When the control voltage to the VCO decreases, the common mode voltage of VCO output increases. Consequently, the leakage current of CML latches inside the CML divider in the hold mode increases, resulting in narrow spikes. The narrow spikes of the CML divider make the PLL feedback frequency larger than expected, causing the narrow tuning range problem. The problem can be fixed by using AC couple or redesigning the CML latch.

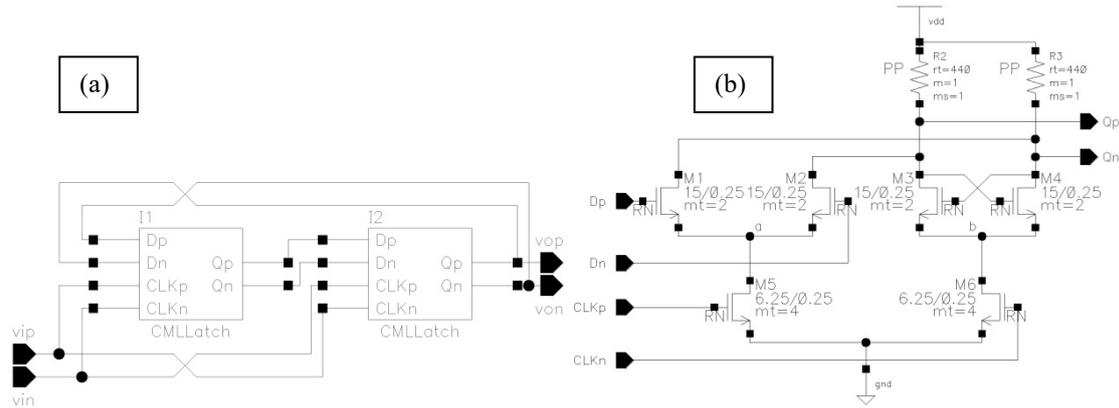

Figure 8. Schematic of the CML divider (a) and the CML latch (b)

## 4. Conclusion

An LCPLL ASIC, fabricated in a commercial 0.25 μm SOS CMOS technology, has been characterized in laboratory test. Random jitter and deterministic jitter are 1.3 ps and 7.5 ps, respectively. The measured tuning range, from 4.6 to 5.0 GHz, is narrower than the expected one, from 3.8 to 5.0 GHz. The narrow tuning range issue has been investigated and understood. The power consumption at the central frequency is 111 mW. After the narrow tuning range problem is fixed, this LCPLL will be used as the clock unit in our next 10 Gbps serializer.

## Acknowledgments

This work is supported by the US-ATLAS R&D program for the upgrade of the LHC and the US Department of Energy grant DE-FG02-04ER41299. The author would like to thank Peter Clarke, Jay Clementson, Yi Kang, Francis M. Rotella, John Sung, and Gary Wu from Peregrine Semiconductor Corporation for technical assistance, Justin Ross at Southern Methodist University for setting up and maintaining the software environment, Jasoslav Ban, Mauro Citterio, Christine Hu, Sachin Junnarkar, Valentino Liberali, Paulo Rodrigues Simoes Moreira, Mitch Newcomer, Quan Sun, Fukun Tang, and Carla Vacchi for technical assistance and reviewing of this design, Hucheng Chen, Francesco Lanni, and Sergio Rescia at Brookhaven National Laboratory for help on varactor measurements.